\documentclass[11pt]{article}
\usepackage{amsmath,amssymb}
\newcounter{appendice}
\newcommand{\appendice}
{
\setcounter{equation}{0}
\renewcommand{\theequation}{\Alph{appendice}.\arabic{equation}}
\addtocounter{appendice}{1}
{\bf Appendix \Alph{appendice}} 
}

\begin{document}

\begin{titlepage}
\title{
{\ }\vspace{-1.2in}\\
{\hfill{\small{SU-4252-745}}}\\\vspace{1.0in}
On the  Failure of Spin-Statistics Connection in Quantum Gravity}

\author{
G. Alexanian \thanks{E-mail address:
garnik@phy.syr.edu}\,\,\,\,\, and \,\,\,\,\,
A.P.Balachandran \thanks{E-mail address:
bal@phy.syr.edu}\\
\\
~
\\
{\small\it 
Physics Department, 
Syracuse University}\\
{\small\it Syracuse NY 13244, USA}\\
} 
\maketitle

\begin{abstract}
Many years ago Friedman and Sorkin 
\cite{Friedman:1980st}
established the existence
of certain topological solitonic excitations in quantum gravity
called (topological) geons. Geons can have quantum numbers 
like charge and can be tensorial or spinorial having
integer or half-odd integer spin. A striking result is that
geons can violate the canonical spin-statistics connection
\cite{Sorkin:1985bg,Aneziris:1989cr}.
Such violation induces novel
physical effects at low energies. The latter will be small
since the geon mass is expected to be of the order of 
Planck mass. Nevertheless, these effects are very striking
and include 
$CPT$ and causality violations and distortion of the cosmic
microwave spectrum. 
Interesting relations of geon dynamics to supersymmetry
are also discussed.

\end{abstract}
\end{titlepage}

\section{Introduction}

The spin-statistics connection asserts that
tensorial particles (those with integral spin)
obey Bose statistics and spinorial ones obey Fermi
statistics. It has a central role in determining
properties of matter including its stability and is 
generally regarded as a fundamental result of 
quantum physics. It has a counterpart in (2+1)
dimensions where particles of fractional spin 
$\theta$ are asserted to obey fractional statistics 
with the same $\theta$.

Our understanding of this connection however is not
perfect. It has been proved using the axioms of
relativistic local quantum field theories (RQFT's)
\cite{Streater:1989vi}.
It has also been established for Skyrme-like
solitons and 'tHooft-Polyakov monopoles
\cite{Sorkin},
and particles of fractional spin in (2+1)
dimensions \cite{Frohlich},
apparently using physical principles which are
mutually different and different too from those
of RQFT's. In the literature, we also encounter
proofs of this connection using yet other
considerations \cite{Berry}. 
In addition, no such theorem can be established
in conventional nonrelativistic physics.
The standard spin-statistics connection cannot also be 
established in generality
for the gravitational topological
excitations known as geons 
\cite{Sorkin:1985bg,Aneziris:1989cr,Balachandran:2001kv,Samuel:1993ua,Dowker:2000zy},
although certain novel spin-statistics connections can be
proved for them \cite{Balachandran:2001kv}.

The lack of spin-statistics connection in quantum 
gravity attracts attention. Nonrelativistic 
physics is a limiting form of RQFT's and therefore
its loss there is attributed to an imperfect 
limiting procedure. But this escape route is not
available for quantum gravity with its
enormous energy scales where RQFT loses
its validity. It is rather the latter which is
a limiting form of a unified model for
gravity and elementary particles.

Studies of the common principles underlying
the different approaches to this connection 
suggest that it needs the possibility of
creation-annihilation processes. Nonrelativistic
models incorporating such events have been devised
(see Balachandran et al. papers in \cite{Aneziris:1989cr}),
they also naturally correlate
spin and statistics.
For geons, these processes can occur only with 
topology change, but even in their presence,
the desired relation can be recovered but imperfectly,
for a limited class of geons \cite{Dowker:2000zy}.
An alternative algebraic approach to quantum gravity
and geon statistics has also been devised with
physical inputs like cluster decomposition \cite{Balachandran:2001kv}.
It predicts definite spin-statistics connection
in (2+1) dimensions [with its probable extension
to higher dimensions], which however does not 
necessarily assert that spinoral geons are
fermions or tensorial geons are bosons.

In summary then, there are strong indications 
that the canonical spin-statistics connection fails
in quantum gravity. We can then enquire how this 
failure percolates to interactions of elementary
particles. We initiate the study of this issue
in this paper. Quantum gravity effects cannot
be important for low energy phenomenology
unless they are enhanced by coherent processes
involving large numbers,  experiments are 
very accurate or proposals of ``large extra dimensions"
\cite{lxd}  are intimations of reality.
But their study is important even if they are tiny
as they challenge concepts of traditional 
quantum physics. If quantum gravity and string physics
are judged by their verifiable predictions, there is no reason
to pursue those enterprises \cite{Woit:2001jy}. 
An added reason for our work
here that it lets us model spin - statistics
violation in a particular way and derive bound on the violation 
parameters.

Geons are discoveries of Friedman and Sorkin
\cite{Friedman:1980st}. Their existence has far-reaching 
implications for quantum gravity.
We begin with a brief introduction to geons and
their spin-statistics properties in Section 2 and
follow it up in Section 3 with the effective interaction
they generate among  (say) standard model particles.
They can be written down using guess work on operator
product expansions, but we go a bit beyond that by identifying
processes that fix their coefficients. The leading
interaction is simple. Geons can be charged or neutral, spinorial 
or tensorial. Let the particle symbols also denote their
fields. A spin-$1/2$ charged geon can then interact with the
electron via the coupling 
${\cal L}^{'}(x)=\eta(e^\dagger G + G^\dagger e)(x)$.
$G$ here is a Bose field: otherwise this interaction 
is not very striking. Similar interactions can happen between 
a tensorial fermion $G$ and standard particles.

An interesting consequence of these interactions is that it can lead
to effective supersymmetry and quantum Hall effect-like phenomena
at low energies.

Interactions of this sort, or more generally even the existence 
of these exotic $G$ fields, are not compatible with RQFT. 
We provisionally take the following stand about this point.
Geons are massive, with mass of the order of Planck mass,
and we look only at low energy processes where they can be handled
nonrelativistically. There is then no inconsistency. At energies
where relativity is important, we presume that new effects enter the
picture, perhaps dictated by the extended structure of geons.
Our models are no good for these energies.

Interactions such as $H^{'}(x)$ have physical
consequences. These are briefly outlined in Sections 4
and 5. In particular, we discuss level distortions
and black body spectrum. There are in addition
violations of
causality and $CPT$ which are also pointed out.
The three Appendices are devoted to technical calculations.

Summarizing, the main results of the paper are 
conceptual and concern the above strikingly novel interaction
between geons and standard elementary particles.
These interactions, studied in the non-relativistic
approximation and low energies here, arise from the
mediation of black holes. They violate 
non-relativistic causality wich requires energy densities at
spatially separated points to commute at a fixed time. They are not 
$CPT$ invariant either. Nevertheless the emergent nonrelativistic 
physics has energy levels bounded below and no obvious inconsistency.
The influences of such an interaction on energy levels and black
body radiation are investigated, but unfortunately no
convincing signal characteristic of the new interaction
and large enough to be detected has been found.
Causality violations can have a sensitive impact on dispersion 
relations \cite{Khuri} and latter can possibly detect the novel 
interactions if the Planck scale is in the TeV. range.
\section{What are Geons}

Elementary approaches to gravity work with spacetimes $X\times \mathbb R$
with $\mathbb R$ accounting for time $t$, and the spatial slice 
$X\times \{t\} \,\approx\, X$
being ${\mathbb R}^D$, except during treatment of black holes.

In the 70's, Friedman and Sorkin \cite{Friedman:1980st} 
initiated studies of asymptotically flat
spatial slices (diffeomorphic to) $X$ different from ${\mathbb R}^D$.
They pointed out that there are classes of manifolds $X$ called prime
manifolds which are perfect infrastructures for describing
elementary solitonic excitations in quantum gravity.
There is only one such  orientable manifold for $D=2$ and that is
the plane with a handle. It leads to the $2D$ geon. A plane with
$n$ handles then gives the excitation of $n$ $2D$ geons.
For $D=3$, there are an infinity of basic manifolds
(connected sums of ${\mathbb R}^3$  and closed prime manifolds)
and an infinity of geons. 
A deep result of Friedman and Sorkin was that quantisation of geons,
 just like the quantisation of two-flavour Skyrmions 
\cite{Balachandran:1991zj}, 
is not unique,
and a certain class of geons can be quantized to give spinorial
particles. The underlying primes are known as spinorial primes.
Don Witt \cite{Witt:1986ef} later extended this work by an exaustive study of 
spinorial primes. Later studies 
\cite{Aneziris:1989cr,Samuel:1993ua} 
revealed that in $2D$,
the geon for a plane with a handle can be quantized to have any spin.

Meanwhile Sorkin \cite{Sorkin:1985bg} studied the spin-statistics
connection for geons and argued that no such relation can exist in
the absence of topology change. This result was elaborated by
Sorkin and coworkers \cite{Aneziris:1989cr} and others
\cite{Balachandran:2001kv,Samuel:1993ua} and
the generic failure of the
spin-statistic connection in quantum gravity was firmly established.
As indicated earlier, a correlation between spin and statistics can be 
shown with enough physical inputs, but still as a rule it fails 
to be convetional.

\section{Effective Interactions Mediated by Geons}

Geons of pure gravity can be tensorial or spinorial,
but will be a singlet under the standard model group. Geons
with nontrivial standard model quantum numbers can occur
when the standard model interactions are also included.
In what follows, we have in mind geons of this enhanced theory
which violate the spin-statistics connection.

We are after novel interactions among elementary particles 
induced by this violation in quantum gravity. Processes leading to such
couplings are not abundant. Those involving black holes
seem to be the sole mediations for this purpose. Black
hole processes conserve quantum numbers like charge and
angular momentum expressible as flux integrals over a sphere 
at infinity. But they need not conserve statistics.
For this reason the following transition can occur. If 
$G$ is a spinorial boson with the same charge as 
the electron $e$,  a black hole can absorb $e$ and emit 
$G$ as Fig.1 illustrates,
and vice versa. This process leads to a direct $e-G$
coupling because of vacuum fluctuations involving the
creation and annihilation of black holes. A virtual black hole
can thus mediate $e-G$ mixing (This process could be 
espesially important should the scenarios presented in
the last two references of \cite{lxd} prove relevant).

Several calculations along these lines exist for gravity-induced
proton decay \cite{Hawking,Adams} where a proton for example is
converted by a black hole into $e$ plus tensorial
particles like photons. Accurate calculations are not possible 
because of lack of control of quantum gravity. The importance of 
such research for the present work is to show that geons will
certainly mix with standard model particles and suggest estimates 
for the coefficients in operator product expansions.

We conclude that black hole fluctuations in the vacuum induce 
$G-e$  couplings with the leading term 
$\lambda e^\dagger G + \lambda^* G^\dagger e$ at low energies.
That is for a charged spin-$\frac{1}{2}$ geon.
A neutral spin-1 geon with real field $G_\mu$ can
likewise couple to the photon field $A_\mu$ by the term 
${\rm constant}\times(\partial_\mu G_\nu - \partial_\nu G_\mu)
(\partial^\mu A^\nu - \partial^\nu A^\mu)$.
We can write other similar quadratic couplings of geons
and low energy 
excitations.
\\
\begin{center}
\begin{picture}(100,80)
\thicklines
\put(50,50){\circle{40}}
\put(10,10){\line(1,1){25}}
\put(10,90){\line(1,-1){25}}
\put(20,20){\line(0,-1){5}}
\put(20,20){\line(-1,0){5}}
\put(20,80){\line(0,-1){5}}
\put(20,80){\line(1,0){5}}
\put(40,85){\makebox(0,0){$G$}}
\put(15,35){\makebox(0,0){$e$}}
\put(50,0){\makebox(0,0){\bf Fig.1}}
\end{picture}
\end{center}

We can assume $\lambda$ to be $>0$ in $G-e$ coupling by writing 
$\lambda=|\lambda|e^{i\chi}$ and absorbing the phase $\chi$ into 
the definition of $G$.
In Fig.1 we can exchange the particles,
so that $\lambda$ has to be a symmetrical function
of $m_e$ and  $m_G$ where  $m_A$ is the mass of particle $A$.
We assume that $m_e<<m_G$ and keep its leading term in
$m_e/m_G$. The symmetry is lost in this approximation.
Dimensional considerations then show that 
$\lambda=m_e \, f(\frac{m_e}{m_G})$
where we retain dependence of $f$ only on the single mass 
ratio $m_e/m_G$, ignoring other elementary particles.
According to \cite{Hawking} (see also \cite{Adams}),
$f\left(\frac{m_e}{m_G}\right)={\left(\frac{m_e}{m_G}\right)}^K$
(times a factor of order 1),
where the integer $K=2\, ({\rm spin\,\,\, of\,\,\, }G)=1$. We let $K$
be free for caution.

The conventional choice for $m_G$ is Planck
mass $m_{pl}\sim 10^{19} Gev$. That gives
$\frac{m_e}{m_G}\sim 10^{-22}$.
In models with ``large extra dimensions'' \cite{lxd},
$m_G$ can be low. For the $Tev$ scale gravity the same ratio
becomes $\frac{m_e}{m_G}\sim 10^{-12}$.

More favorable values of $\lambda$ can be got
by changing $e$ to a heavier particle. Already with a 
neutron we gain a factor of $10^3$:
\begin{eqnarray}
m_n/m_G &\approx& 10^{-19}\ \ \ {\rm if \ } m_G\sim m_{Pl},\\
{}&\approx& 10^{-9}\ \ \ {\rm if \  } m_G\sim 1\ \ {Tev}.\nonumber
\end{eqnarray}
In this case, if $K=1$ or $2$, the effects studied below
are within experimental reach \cite{Greenberg,Okun}.

\section{Level Distortions} 

In this section, we explore the effects of the interaction 
in section 3 on energy levels. They get shifted as is to be
expected.
This effect is illustrated using 
the harmonic oscillator system. It is a simple, but basic
system where the new physics can be understood with
relative transparency and then applied to other
situations. It is also an approximation to quantum field theory
where we retain only one mode each of a geon and a standard model
field and only terms in the Lagrangian density quadratic in
these fields 
(see below).

{\it i) A Boson  and a  Fermion} 

There are new important features encountered when more 
than one fermion or boson is considered in the Hamiltonian.
We will therefore study them later.

%
Let us consider a system that has only two degrees of freedom,
represented by creation operators $(b^\dagger, f^\dagger)$
an annihilation operators $(b,f)$. 
Commutation (anticommutation) relations are taken to be
\begin{equation}
[b,b^\dagger]=1,\ \ \ \ \ \{f,f^\dagger\}=1
\label{comm} 
\end{equation}
where curly brackets mean anti-commutators as usual.
Assuming the existence of the common vacuum, $|0\rangle$,
which is annihilated by both $f$ and $b$, 
the Hilbert space is spanned by the linear combinations of the
following states:
\begin{eqnarray}
&&|n\rangle=\frac{1}{\sqrt{n!}}(b^\dagger)^n\,|0\rangle, \ \ \ \
f^\dagger|n\rangle=\frac{1}{\sqrt{n!}}(b^\dagger)^n
\,f^\dagger|0\rangle\\
&&b|0\rangle=f|0\rangle=0\nonumber
\end{eqnarray}

As has already been mentioned in the introduction,
a relativistic
particle with half-integer (integer) spin can only  be 
successfully described in a {\it local} field 
theoretic formalism if it has fermionic (bosonic) commutation 
rules. Since the excitations that we want to study are
very heavy ($m\approx M_{pl}$), 
they should admit a non-relativistic description.
In this case there is no connection between spin and statistics
and we will therefore assume that one of the operators
(it does not matter for the moment which one) represents the 
excitation with the ``wrong'' statistics
(e.g. either $b$'s are spin half
or $f$'s are spin 0,1 etc.). 
[The spin degrees of freedom are being ignored.]

At this point, a few comments are in order.
The Hamiltonian of the model is
\begin{equation}
H=\omega_{b}b^\dagger b + \omega_{f}f^\dagger f + g b^\dagger f +
g f^\dagger b
\label{hamiltonian} 
\end{equation}
where we can assume that $g>0$ 
as pointed out earlier.

We can obtain (\ref{hamiltonian}) for example from
a Hamiltonian density $\cal H$ with standard free field terms
for $G$ and $e$ and an additional interaction ${\cal H}^{'}:$
\begin{eqnarray}
{\cal H}&=&-\frac{1}{2m}G^\dagger\,\nabla^2 G - e^\dagger
(\vec\alpha\cdot\vec p + \beta\,m_e)\,e + {\cal H}^{'},\nonumber\\
{\cal H}^{'}&=&\eta(G^\dagger\,e+e^\dagger\,G).\nonumber
\end{eqnarray}

To get (\ref{hamiltonian}) we then mode expand $G$ and $e$
so as to diagonalize the free field terms. On retaining
just one mode in these expansions (pretending that they are
discrete) and including ${\cal H}^{'}$, we get (\ref{hamiltonian}).

Below we will diagonalize (\ref{hamiltonian}).
The {\bf generic} eigenstates of $H$ are not created from the vacuum
by simple linear expressions in $b^\dagger$ and $f^\dagger$ and their 
powers. Rather they are created from the vacuum by complicated 
expressions involving $b^\dagger$ and $f^\dagger$. For this reason, the mode 
expansion diagonalizing the Hamiltonian 
$\int d^3x\, {\cal H}(x)$ is unknown to us.

The problem can be seen in yet another manner.
The Hamiltonian for the $G$,$e$ fields has the form
\begin{eqnarray}
\int\!\!\!&d^3x&\!\!\!
 \left(G^\dagger+e^\dagger\right)(x)\,{\hat {\cal H}}\,
 \left(\begin{array}{c}
	G\\
	e
	\end{array}
 \right) (x),
\nonumber\\
{\hat {\cal H}}\!\!\!&=&\!\!\!\left(\begin{array}{cc}
-\frac{1}{2m}\nabla^2&\eta\\
\eta&\vec\alpha\cdot\vec p + \beta\,m_e
\end{array}\right)
\nonumber
\end{eqnarray}
where $G$ and $e$ are say Bose and Fermi fields and $\hat {\cal H}$
is the ``single particle Hamiltonian''. Normally
we would expand $G$ and $e$ in terms of eigenstates of $\hat {\cal H}$
with creation and annihilation operators of appropriate
statistics as coefficients. But that does not work now. 
The eigenstates are given by $\hat {\cal H}\Psi_n=E_n \Psi_n$
and have the form
\begin{equation}
\Psi_n=\left(
\begin{array}{c}
  \beta_n\\\phi_n
 \end{array}
\right).
\nonumber
\end{equation}
They mix Bose and Fermi modes, $(\beta_n,0)$ and $(0,\phi_n)$
not being eigenstates. But then, we do not know what statistics 
to assign to $a_n$ in the expansion
\begin{equation}
\left(
\begin{array}{c}
  G\\e
 \end{array}
\right)=\sum\,a_n\,\Psi_n.
\nonumber
\end{equation}
Let us return to study of the spectrum of this 
Hamiltonian.
It is easy to construct the exact eigenstates of this model
(see Appendix A).
The Schr\"odinger equation is easily solved by
using the ansatz 
\begin{eqnarray}
|\psi\rangle_n = (\alpha_n(b^\dagger)f^\dagger+\phi_n(b^\dagger) )|0\rangle\\
\label{state} 
H|\psi\rangle_n=E_n|\psi\rangle_n
\end{eqnarray}
The spectrum is given by two series of states labeled by 
the non-negative integer $n$
with $\alpha_n(x)\sim x^n$ and $\phi_n(x)\sim x^{n+1}$ (plus the vacuum 
state, $|0\rangle$ which remains an eigenstate even for
$g\ne 0$). Energies of the pair of $n^{th}$ states 
are (see (\ref{spectrum1b1f}))
\begin{equation}
E_n^{\pm}=\frac{1}{2}\left(\omega_b(2n+1)+\omega_f \pm \sqrt{(\omega_b-\omega_f)^2
+4g^2(n+1)}\right).
\label{energy}
\end{equation}
This is the complete spectrum of the system. As $g$ tends to 0
each state smoothly goes into one eigenstate of the 
unperturbed Hamiltoinian.
It is interesting that perturbative ground state may or may not remain
as such depending on the value of $g$.
The condition 
for the existence of
negative eigenvalues in the spectrum is
\begin{equation}
n\omega_b^2+\omega_b\omega_f<|g|^2
\label{condition}
\end{equation}
The minimal value of  $|g|^2$ when this happens is therefore
$|g|=\sqrt{\omega_b\omega_f}$ while the approximate number of
negative eigenvalues is $n_{-}=(|g|^2-\omega_b\omega_f)/\omega_b^2$.
It is worth mentioning that a similar Hamiltonian with
both $f$ and $b$ considered to be bosons
will $not$ have a ground state once $g$ is sufficiently large
($|g|>\sqrt{\omega_b\omega_f}$).

A striking feature of these levels is that they are $b-f$ mixtures.
For a small $g$ and $\omega_f>\omega_b$ we find the following behaviour:
\begin{eqnarray}
\alpha^{+}_n&=&\left(1-\frac{g^2\,(n+1)}{2(\omega_b-\omega_f)^2}
+\cdots\right)\frac{1}{\sqrt{n!}}
\\
\phi^{+}_n&=&\frac{g}{(\omega_f-\omega_b)}\frac{1}{\sqrt{n!}}
\label{coeff_att_small_g}
\\
\alpha^{-}_n&=&-\frac{g}{(\omega_f-\omega_b)}\frac{\sqrt{n+1}}{\sqrt{n!}}
\\
\phi^{-}_n&=&\left(1-\frac{g^2\,(n+1)}{2(\omega_b-\omega_f)^2}
+\cdots\right)\frac{1}{\sqrt{(n+1)!}}
\end{eqnarray}
(In case of $\omega_f<\omega_b$, one 
makes the interchanges
$\alpha^+_n\leftrightarrow\alpha^-_n,\ \ \phi^+_n\leftrightarrow\phi^-_n$).

A point worthy of attention is that level degeneracy for
the generic level 
is not 
affected as $g$ is turned on (with the exception of the occasional
coincidence of energies for some special values of $g$).
As it is increased from 
zero, each level adiabatically and smoothly evolves.
Degeneracy of the levels is therefore not affected
when $g$ becomes non-zero.
  
However, there is one special case $g=g_s=\sqrt{\omega_b\omega_f}$ 
which makes
(\ref{condition}) into an equality. 
In this case $E^-_0=0$ as it is an eigenvalue for the
perturbative vacuum $|0\rangle$ (remember that it is 
still an eigenstate). So in this case the ``ground'' state
becomes degenerate, with two states of the same energy
$0$ being
\begin{eqnarray}
|0\rangle,\ \ \ 
\left(-\sqrt{\frac{\omega_b}{\omega_f+\omega_b}}f^\dagger
+\sqrt{\frac{\omega_f}{\omega_f+\omega_b}}b^\dagger 
\right) |0\rangle.
\end{eqnarray}   

{\it ii) Connection to Supersymmetry and QHE.}

It has been pointed out to us by Joseph Samuel that the
Hamiltonian (\ref{hamiltonian}) is an element of the 
graded algebra
of a supergroup, with $SO(2)$ as its underlying classical
group. One sees this from the anticommutator
\begin{equation}
\left[ b^\dagger\,f,f^\dagger\,b\right]_+=b^\dagger\,b+ f^\dagger\,f.
\end{equation}
The graded algebra has $b^\dagger\,b$ and $f^\dagger\,f$ as 
even generators and $b^\dagger\,f$ and its adjoint as odd generators.

There is also  an interesting connection of (\ref{hamiltonian})
to the Dirac Hamiltonian in a plane with a perpendicular
uniform magnetic field as has also been
pointed out to us by Samuel. In this case 
the three-dimensional zero-mass Dirac Hamiltonian in a 
magnetic field, $\vec\alpha\cdot\vec\pi$, becomes 
$\alpha^+\pi^-+\alpha^-\pi^+$, $\{\alpha^+,\alpha^-\}=1,[\pi^+,\pi^-]=eB$
where $B$ is the magnetic field along the perpendicular direction.
This Hamiltonian 
can be identified with the last two terms in (\ref{hamiltonian}).  

{\it iii) One Boson and Two Fermions.}

In the hydrogen atom, there are two levels with principal 
quantum number $n=1$ corresponding to spin up and spin down.
Their creation operators $f_i^\dagger\ \ \ (i=1,2)$ in the 
second-quantized formalism anticommute. Similarly we can 
associate fermionic oscillators to bound state levels of 
any spinorial particle.

This association is in conventional physics in the absence
of the disturbing presence of geons.
With geons in the spectrum there are additional
interactions which spoil such simple associations.
The simpliest model that we can consider is a  generalization 
of the Hamiltonian (\ref{hamiltonian}) to the two
fermionic modes $f_{1,2}$ interacting with a single 
bosonic geon $b$:
\begin{eqnarray}
H=\omega_{1}f_{1}^\dagger f_{1} + \omega_{2}f_{2}^\dagger f_{2}
+\Omega b^\dagger b + g_{1}(f_{1}^\dagger b +
b^\dagger f_{1})+ g_{2}(f_{2}^\dagger b +
b^\dagger f_{2})
\label{twofermions}
\end{eqnarray}
This Hamiltonian is analyzed in Appendix B.
One can show that for the most interesting case
when both fermions are {\it degenerate}, i.e
$\omega_1=\omega_2$ it is possible to find the
eigenfunctions and exact spectrum of the model.
We shall show that the operator for the level
corresponding to two fermionic excitations has the form
(\ref{twofassumption}):
\begin{eqnarray}
|\psi\rangle=\left\{\alpha(b^\dagger)f^\dagger_1 f^\dagger_2
+ \phi_1(b^\dagger)f^\dagger_1+\phi_2(b^\dagger)f^\dagger_2+
\Psi(b^\dagger)\right\}|0\rangle
\end{eqnarray}
It goes over to just $f^\dagger_1\,f^\dagger_2$ as the interaction
with the geon is switched off. 

Let us next consider the case where $\omega_1=\omega_2$
and $g_1=g_2=g$. In this case, (\ref{twofermions})
is invariant under the exchange of $f_1$ with $f_2$.
In order to study the possible Pauli principle violation
in this case, 
one should consider what happens when the two
fermionic operators are exchanged. Then $|\psi\rangle$
becomes $|\psi^{'}\rangle$, where
\begin{eqnarray}
|\psi'\rangle=\left\{\alpha(b^\dagger)f^\dagger_2 f^\dagger_1
+ \phi_1(b^\dagger)f^\dagger_2+\phi_2(b^\dagger)f^\dagger_1+
\Psi(b^\dagger)\right\}|0\rangle.
\end{eqnarray}
Thus $|\psi\rangle$
will be an eigenstate of the permutation
operator with -1 eigenvalue only if $\Psi=0$ and $\phi_1=-\phi_2$.
In general, if $\omega_1\ne\omega_2$ and/or $g_1\ne g_2$, 
this is not the case. However,
we have checked that for $\omega_1=\omega_2$ and $g_1=g_2$,
for the eigenvalue series that goes to $E_3=2\omega+n\,\Omega$
in the limit $g_{1,2}\rightarrow 0$, this is precisely
the case: the eigenvectors of the (\ref{twofmatrix}) matrix have 
the structure
$(X_n(b),Y_n(b),-Y_n(b),0)$.
As we expect this result
to be generally true, we conclude that there is no 
apparent Pauli principle violation
in this model.

Actually we can argue that models like this cannot violate
Pauli principle unless level degeneracy is affected as 
$g(=g_1=g_2)$ becomes nonzero. That is because
(\ref{twofermions}) for $\omega_1=\omega_2$ and
$g_1=g_2=g$ is symmetric under exchange of $f_1$ and $f_2$
and hence its eigenstates can be organized in irreducible
representations (IRR's) of the permutation group $S_2$.
At $g=0$, the energy eigenstates with two fermions 
change sign under their exchange: they transform by
the nontrivial IRR of $S_2$. By continuity, this IRR will
persist if $g$ is made nonzero. New effects can arise 
if level degeneracy is changed when $g$ becomes nonzero,
so that there is a state symmetric under $S_2$ degenerate 
with this IRR. But that does not happen in our model.

\section{Black Body Spectrum}

As it is written now, either of the 
frequencies $\omega_{b,f}$ may be taken to 
correspond to a geon, the actual choice is dictated by
the low energy effect that one wants to study. In the case of  
effects in atomic systems,
one assumes
that $\omega_b\sim M_{pl}$ and $\omega_f=\sqrt{m^2_e+k^2}$,
$\omega_f<<\omega_b$. In this case the geon
is a spin half bosonic excitation in gravity.
From the cosmological point of view,
however, it is very interesting to look at the case
%
of the microwave background radiation, where the geon
must be the spin 1 excitation and 
hierarchy of scales is opposite:  
$\omega_b\sim k_{photon}$ and $\omega_b<<\omega_f\sim M_{pl}$.


In order to determine the effect of geons on the Planck spectrum
one has to determine the correction to the thermal distribution function
for the photons. Since the spectrum of the Hamiltonian 
(\ref{energy}) is known, the task reduces to the computation of the
proper partition function (see Appendix C for details), which can be
done perturbatively in $|g|^2$.
The first non-trivial correction to the distribution function $n$
turns out to be given by (\ref{distribution1}),
\begin{eqnarray}
n(\omega)=n_0(\omega)-n_0(\omega)\frac{|g|^2}{M_{Pl}^2}+
\frac{e^{\beta\omega}}{(e^{\beta\omega}-1)^2}{\beta|g|^2\over M_{Pl}}
+\cdots\ ,\ \ \ \ n_0(\omega)=\frac{1}{e^{\beta\;\omega}-1}
\label{hhh}
\end{eqnarray}
where $n_0$ is the free bosonic Planck distribution.

One can immediately see that the expansion of (\ref{hhh})
in powers of $|g|^2$ is
singular as $\omega_b$ goes to $0$: the first correction diverges
as $1/\omega_b^2$. This is the same problem that plagues any theory
that has massless modes at non-zero temperature.
However, in this case a more careful treatment is needed.
The reason for that is that the proliferation of the soft
photons in the usual, free case does not lead to a divergent
energy density: the phase space volume scales like $\omega^3\,d\omega$
and the denominator of the bosonic distribution produces a $1/\omega$
factor thus keeping the product finite.
For us, the next order correction produces terms
like $1/\omega^k$ where $k$ is roughly proportional
to the order of the perturbative expansion.
One possible solution to this problem is
the following: in the case of the photon,
due to dimentional considerations one should consider
$g\sim k^2/M_{pl}$, which will keep the answer finite
in the limit $k=\omega_b\rightarrow 0$.
This behaviour of $g$ is supported by the fact that the
leading gauge invariant geon-photon coupling involves 
derivatives being $constant\times(\partial_\mu A_\nu-\partial_\nu A_\mu)
(\partial_\mu G_\nu-\partial_\nu G_\mu)$ for a spin 1 geon 
as indicated earlier.
Numerically, one can stop at the first order
if $(\beta\omega_b)^2>>\beta|g|^2/(\omega_f)$. Using
the suggestion above this condition reduces to just
$T<<M_{pl}$ which is obviously satified.


The first correction to $n_0$ in the formula (\ref{hhh}) is just a ``grey
body'' factor, while the second one is more important at low
frequencies of the photon. However, at this point one has to use
the expression $\sim k^2/M_{Pl}$ for $g$, which makes the 
corrections look like
\begin{eqnarray}
\Delta n(\omega) = -n_0(\omega)\frac{k^4}{M_{Pl}^4}+
\frac{e^{\beta\omega}}{(e^{\beta\omega}-1)^2}
{\beta k^4\over M_{Pl}^3}
\end{eqnarray}
In the limit $k=\omega\rightarrow 0$ the second term becomes
$k^2/(\beta M_{Pl}^3)$ and it is clearly insignificant
for small $k$.
It seems then, that even though there are some corrections to the
``background radiation'' following from the
Hamiltonian described above they are too tiny to be detected
in the experimentally accessible region. 

It is important to make sure that whatever correction
that the observed distribution function gets is a signature of
the effect in hand and not of some other origin.
While it is almost certainly impossible to establish
this rigorously, one elementary test is
possible here - what if the ``mixed'' mode is a boson as well?
Now, for comparison let us consider the similar situation in the case when
we have two bosonic operators coupled the same way as in the Hamiltonian
(\ref{hamiltonian}):
\begin{equation}
H=\omega_{1}a^\dagger_1 a_1 + \omega_{2}a_2^\dagger a_2 + g a_1^\dagger a_2 +
g^* a^\dagger_2 a_1
\end{equation}
Here both $a_1$ and $a_2$ are bosons.

The correction to order $|g|^2$ to 
the distribution function 
can be calculated from (\ref{bosonboson}),
the result is identical to that of
(\ref{distribution1}) to order 
$|g|^2$, with the same ``grey body'' factor
and correction to the low energy behaviour.
Thus, unfortunately, the difference
between (\ref{distribution1}) and (\ref{bosonboson}), therefore, 
comes only in the next order of perturbation theory. 
While the full expressions are given by (\ref{secondorder}),
this difference in powers of $|g|^2$ is  
\begin{eqnarray}
\Delta n(\omega)_{b+g}-
\Delta n(\omega)_{b+b}=\left\{
\frac{\beta}{M^3_{Pl}}\left(\frac{2-6e^{\beta\omega}}
{(e^{\beta\omega}-1)^3}\right)+
\frac{6}{M^4_{Pl}}\left(\frac{e^{\beta\omega}}{(e^{\beta\omega}-1)^2}\right)
\right\}g^4+\cdots
%
\end{eqnarray}
where $\Delta n(\omega)_{b+f}$ is $n(\omega)-n_0(\omega)$ in (\ref{hhh}),
while $\Delta n(\omega)_{b+b}$ is given in (\ref{secondorder}).
These corrections become identical in the ultraviolet limit
$\beta\omega\rightarrow \infty$ and the only difference
comes in the subleading order in $1/M_{Pl}$.
For that reason, it is difficult to propose an experimental test
which would be able to see these corrections.

\section{Final remarks: $CPT$, Causality}

{\it i) CPT}.

In the course of proving the $CPT$ theorem, 
anti-commutativity of fermionic fields and 
commutativity of tensorial ones
are explicitly used \cite{Streater:1989vi}. 
But this feature need not hold for geons. 
$CPT$ thus can fail in the presence of
geons. The failure will be by small numbers like 
$10^{-19}$,$10^{-9}$ or its powers [cf.Eq. (1)]. 
Detailed calculations may be possible by allowing for mixing 
of quarks and leptons for instance with geons and integrating out
the latter, but we have not done this work.

{\it ii)  Causality}.

If $\Psi$ is a spinorial fermion field and $G$ a spinorial
boson field, for example, the term 
${\cal H}_I(x)=\lambda(G^\dagger\Psi+\Psi^\dagger G)(x)$ in the
Hamiltonian density ${\cal H}(x)$ does not commute for the
space-like separations: $[{\cal H}(x),{\cal H}(y)]\ne 0$, $(x-y)^2>0$.
As ${\cal H}(x)={\cal H}_0(x)+{\cal H}_I(x)$ where ${\cal H}_0(x)$ 
commutes for spacelike separations, ${\cal H}(x)$ and ${\cal H}(y)$
neither commute nor anti-commute for space-like 
separations. Thus Hamiltonian density, an observable, 
violates local causality.

Our model is in reality valid only nonrelativistically,
so that we have to interpret this statement as asserting
that the Hamiltonian densities at distinct spatial points 
at the same time do not commute. (Hence they cannot be 
``simultaneously measured''.)

The implications of this microscopic violation of causality are
not adequately clear to us. It does have a phenomenological
implication:
forward dispersion relations will not be correct.
Such violations of causality also occur in noncommutative geometry, 
in particular of D-branes in string physics \cite{Sheik}.
Basically, this causality violation in our 
model is controlled by
the intrinsic
non-locality of the geons, and this non-locality 
is similar to having a 
fundamental length $l_f\sim 1/M_{Pl}$ in the theory.
There are some indications \cite{Khuri} that
forward dispersion relations can be a sensitive probe
of $l_f$ provided it is not too small, say if $1/l_f$
is in the TeV range.
Nonlocality will also spoil the analyticity of scattering amplitudes
and its implications by
small corrections. Investigations of these effects would be of great 
interest, being characteristic manifestations of intrusions of
quantum gravity or string physics into elementary particle theory.

\section{Acknowledgements}

We have benefited from much help, advice and 
crucial observations from
Denjoe O'Connor, Joseph Samuel, Ajit Srivastava and Rafael Sorkin.
This work was supported by DOE and NSF under contract numbers
DE-FG02-85ER40231 and INT-9908763 respectively.



\appendice

In this appendix we construct eigenstates and find eigenvalues
for the Hamiltonian (\ref{hamiltonian}). 
We start by writing an arbitrary eigenstate $|\psi\rangle$ as
\begin{eqnarray}
|\psi\rangle &=& (\alpha(b^\dagger)f^\dagger+\phi(b^\dagger) )|0\rangle,\\
H|\psi\rangle&=&E|\psi\rangle
\label{shreq}
\end{eqnarray}
Using commutation relations (\ref{comm}) one has
\begin{eqnarray}
\left[b,\phi(b^\dagger)\right]=
\frac{\partial \phi(b^\dagger)}{\partial b^\dagger},\ \ \ \ 
\left[b,\alpha(b^\dagger)\right]=
\frac{\partial \alpha(b^\dagger)}{\partial b^\dagger}
\label{comm_with_b}
\end{eqnarray}
and equation (\ref{shreq}) becomes
\begin{eqnarray}
\left\{
\omega_f\alpha(b^\dagger)\,f^\dagger+
\omega_b\frac{\partial\alpha(b^\dagger)}{\partial b^\dagger}\,b^\dagger
\,f^\dagger+
g\,\alpha(b^\dagger)\,b^\dagger+
\omega_b\frac{\partial\phi(b^\dagger)}{\partial b^\dagger}\,b^\dagger+
g\,\frac{\partial\phi(b^\dagger)}{\partial b^\dagger}\,f^\dagger
\right\}|0\rangle=\nonumber\\
=E\,\left\{\alpha(b^\dagger)\,f^\dagger+\phi(b^\dagger)\right\}
|0\rangle.
\end{eqnarray}
By rewriting this in the matrix form one gets:
\begin{eqnarray}
\omega_f\alpha(b^\dagger)+
\omega_b\frac{\partial\alpha(b^\dagger)}{\partial b^\dagger}\,b^\dagger+
g\,\frac{\partial\phi(b^\dagger)}{\partial b^\dagger}&=& 
E\,\alpha(b^\dagger),\\
g\,\alpha(b^\dagger)\,b^\dagger+
\omega_b\frac{\partial\phi(b^\dagger)}{\partial b^\dagger}\,b^\dagger&=&
E\,\phi(b^\dagger).
\label{shreq2}
\end{eqnarray}
At this point let us assume the following behaviour of the
functions $\alpha$ and $\phi$:
\begin{eqnarray}
\alpha(b^\dagger)=\alpha_n\,(b^\dagger)^{n},\ \ \ \
\phi(b^\dagger)=\phi_n\,(b^\dagger)^{(n+1)}.
\end{eqnarray}
Here $\alpha_n$ and $\phi_n$ are (complex) numbers.
It is easy to see that with this choice equations
(\ref{shreq2}) reduce to a simple eigenvalue problem
for a $2 \times 2$ matrix:
\begin{eqnarray}
\left(\begin{array}{cc}
\omega_f+n\,\omega_b & g\,(n+1).\\
g & (n+1)\omega_b
\end{array}
\right)\,\left(\begin{array}{c}\alpha_n\\ \phi_n\end{array}\right)
=E_n\,\left(\begin{array}{c}\alpha_n\\ \phi_n\end{array}\right)
\label{matrix2}
\end{eqnarray}
The eigenvalue equation is quadratic,
and yeilds the following two sets of solutions:
\begin{equation}
E_n^{\pm}=\frac{1}{2}\left(\omega_b(2n+1)+\omega_f \pm \sqrt{(\omega_b-\omega_f)^2
+4g^2(n+1)}\right).
\label{spectrum1b1f}
\end{equation}
With these values of energy the coeficients $\alpha_n$ and $\phi_n$
can be determined from the normalization condition 
$n!|\alpha_n|^2+(n+1)!|\phi_n|^2=1$
and equation (\ref{matrix2}).
They are found
(upto an over-all phase) to be
\begin{eqnarray}
\alpha^\pm_n&=&\frac{\Delta\omega\pm\sqrt{\Delta\omega^2+4g^2(n+1)}}
{\left(\left(\Delta\omega\pm\sqrt{\Delta\omega^2+4g^2(n+1)}\right)^2
+4g^2(n+1)\right)^{1/2}}\frac{1}{\sqrt{n!}}\\
\phi^\pm_n&=&\frac{2g}
{\left(\left(\Delta\omega\pm\sqrt{\Delta\omega^2+4g^2(n+1)}\right)^2
+4g^2(n+1)\right)^{1/2}}\frac{1}{\sqrt{n!}}\\
\end{eqnarray}
where $\Delta\omega=\omega_f-\omega_b$.
These expressions characterize the complete spectrum of the model.
In the limit $g\rightarrow 0$, this 
smoothly goes to the unperturbed 
spectrum for which $g=0$. This ensures that 
we have found all of the  eigenstates of the system.


\appendice

A similar treatment can be applied to the case of two
fermionic modes coupled to a boson as in the
Hamiltonian (\ref{twofermions}).
The most generic ansatz for an energy eigenstate is 
\begin{equation}
|\psi\rangle=\left\{\alpha(b^\dagger)f^\dagger_1 f^\dagger_2
+ \phi_1(b^\dagger)f^\dagger_1+\phi_2(b^\dagger)f^\dagger_2+
\Psi(b^\dagger)\right\}|0\rangle.
\end{equation}
Applying Hamiltonian (\ref{twofermions}) and using
relations similar to those of (\ref{shreq})
and (\ref{comm_with_b}),
we get
\begin{eqnarray}
(\omega_1+\omega_2)\alpha(b^\dagger)+
\Omega_b\frac{\partial\alpha(b^\dagger)}{\partial b^\dagger}\,b^\dagger+
-g_2\frac{\partial\phi_1(b^\dagger)}{\partial b^\dagger}+
+g_1\frac{\partial\phi_2(b^\dagger)}{\partial b^\dagger}=
E\,\alpha(b^\dagger)
,\nonumber\\
-g_2\alpha(b^\dagger)b^\dagger+
\omega_1\phi_1(b^\dagger)+
\Omega\frac{\partial\phi_1(b^\dagger)}{\partial b^\dagger}b^\dagger+
g_1\frac{\partial\Psi(b^\dagger)}{\partial b^\dagger}=E\phi_1(b^\dagger)
,\nonumber\\
+g_1\alpha(b^\dagger)b^\dagger+
\omega_2\phi_2(b^\dagger)+
\Omega\frac{\partial\phi_2(b^\dagger)}{\partial b^\dagger}b^\dagger+
g_2\frac{\partial\Psi(b^\dagger)}{\partial b^\dagger}=E\phi_2(b^\dagger)
,\nonumber\\
g_1 \phi_1(b^\dagger) b^\dagger + g_2 \phi_2(b^\dagger) b^\dagger+
\Omega\frac{\partial\Psi(b^\dagger)}{\partial b^\dagger}b^\dagger
=E\Psi(b^\dagger).
\label{twofeq}
\end{eqnarray}
As in the previous case we first assume the
power-law behaviour of coefficients $\alpha,\phi_{1,2},\Psi$
and write
\begin{eqnarray}
\alpha(b^\dagger)_n=\alpha_n(b^\dagger)^n,\  
\phi_1(b^\dagger)_n={\phi_1}_n(b^\dagger)^{n+1},\  
\phi_2(b^\dagger)_n={\phi_2}_n(b^\dagger)^{n+1},\  
\Psi(b^\dagger)_n=\Psi_n(b^\dagger)^{n+2}.
\label{twofassumption}
\end{eqnarray}
The corresponding ($4\times 4$) matrix equation
can be read off the equations (\ref{twofeq}):
\begin{equation}
\left(
 \begin{array}{cccc}
   \omega_1+ \omega_2+\Omega n & -g_2(n+1) & g_1(n+1) & 0\\
   -g_2  &   \omega_1+\Omega(n+1)& 0 & g_1(n+2)\\
   +g_1  & 0 &  \omega_2+\Omega(n+1) & g_2(n+2)\\
      0  & g_1 & g_2 & \Omega(n+2)
 \end{array}
\right)\,
\left(
 \begin{array}{c}
   \alpha_n\\{\phi_1}_n\\{\phi_2}_n\\ \Psi_n
  \end{array}
\right)
=E\,
\left(
 \begin{array}{c}
   \alpha_n\\{\phi_1}_n\\{\phi_2}_n\\ \Psi_n
  \end{array}
\right)
\label{twofmatrix}
\end{equation}
The eigenvalues can be obtained from the fourth order
secular equation and are quite complicated for
generic values of the frequencies and coupling constants.
Nevertheless for the physically interesting case
of  degenerate ($\omega_1=\omega_2=\omega$)
fermions coupled to a bosonic ``geon'', the 
eigenvalue equation for the matrix (\ref{twofmatrix})
splits into the product of two quadratic ones
and yields
\begin{eqnarray}
(\omega+\Omega(n+1) - E)(\Omega(n+2)-E)-(g_1^2+g_2^2)(n+2)=0,\\
(2\omega+\Omega n - E)(\omega+\Omega(n+1)-E)-(g_1^2+g_2^2)(n+1)=0.
\end{eqnarray}
The corresponding energy eigenvalues are
\begin{eqnarray}
{E}_{1,2}&=&\frac{1}{2}\left\{
\omega+\Omega(2n+3)\pm\sqrt{
(\omega-\Omega)^2+4\,g^2\,(n+2)}\right\}\nonumber\\
{E}_{3,4}&=&\frac{1}{2}\left\{
3\omega+\Omega(2n+1)\pm\sqrt{
(\omega-\Omega)^2+4\,g^2\,(n+1)}\right\}\\
g^2&=&g_1^2+g_2^2\nonumber
\end{eqnarray}

But these do not exhaust all the energy eigenstates.
Thus, if we look at the limiting case of $g_{1,2}\rightarrow 0$,
we find 
\begin{eqnarray}
{E}_{1}=\omega+\Omega(n+1),\ \ \ 
{E}_{2}=\Omega(n+2),\nonumber\\
{E}_{3}=2\omega+\Omega n,\ \ \ 
{E}_{4}=\omega+\Omega(n+1).
\end{eqnarray}
Taking into account that vacuum state $|0\rangle$
remains an eigenstate with energy $0$, we see that
there are three eigenvalues that are missing in the above
sets, namely $\Omega,\omega_1,\omega_2$.
This has happened because while solving
(\ref{twofeq}) we assumed  (\ref{twofassumption}), which is not
the only possibility.
Assuming that $\alpha(b^\dagger)=0$, one can show that
the following equations for $\phi_1,\phi_2$ and $\Psi$
result:
\begin{eqnarray}
g_2\phi_1(b^\dagger)=g_1\phi_2(b^\dagger).\nonumber
\end{eqnarray}
These equations bring in the ``missing'' energies.


\appendice

In this  appendix we give detailed calculations of the
influence of the boson-fermion mixing on the black body
radiation spectrum (microwave background).

The standard way to do so is to introduce
chemical potentials   $\mu_f, \mu_b$  for fermionic and bosonic excitations
of the model.
Then
\begin{equation}
n_b(\omega_b)=-\frac{\partial\Omega}{\partial \mu_b}|_{\mu_b=0}
\end{equation}
where $n_A(\omega_A)$ is the mean number of particles of type A
with energy $\omega_A$, and the 
potential $\Omega$ is given by 
\begin{equation}
\Omega=-\frac{1}{\beta}\,{\rm ln}\, Z,\ \ \ \ Z={\rm tr}\,e^{-\beta(H-\mu_b
a^\dagger a -\mu_f b^\dagger b)}.
\end{equation}
Because our initial $H$ is quadratic, the addition of number operators
leads just to the effective changes $\omega_b\rightarrow \omega_b-\mu_b$
and $\omega_f\rightarrow \omega_b-\mu_f$. Since the trace can be computed
over any set of complete states, one can use these new values
in the expression for the spectrum (\ref{energy}) and just sum over $n$:
\begin{eqnarray}
&&Z=1+ \sum_{n=0}^{\infty}\{e^{-\beta E^+_n}+e^{-\beta E^-_n}\}
\\
&&= 1+2\sum_{n=0}^{\infty}e^{-\frac{\beta}{2}((2n+1)\omega_b+\omega_f)}
{\rm ch}\left(\frac{\beta}{2}
\sqrt{(\omega_b-\omega_f)^2+4|g|^2(n+1)}\right)\nonumber
\end{eqnarray}
As the coupling parameter $g$ goes to $0$,
the above expression tends to
\begin{eqnarray}
&\ &1+\sum_{n=0}^{\infty}\{e^{-\beta(n+1)\omega_b}+e^{-\beta (n\omega_b+\omega_f)}\}
\nonumber\\
&=&\frac{1+e^{-\beta\omega_f}}{1-e^{-\beta\omega_b}}=Z_f\times Z_b
\end{eqnarray}
where $Z_f$ and $Z_b$ are free fermionic and free bosonic partition
functions.
Expanding in powers of $|g|^2$, one gets
\begin{eqnarray}
Z&=&Z_f\,Z_b+2\,\frac{e^{-\frac{
\beta}{2}(\omega_f+\omega_b)}}{(1-e^{-\beta\omega_b})^2}
\frac{
      \beta\,{\rm sh}\left(-\frac{\beta}{2}|\omega_b-\omega_f|\right)
     }{|\omega_b-\omega_f|}
|g|^2+\cdots.
\end{eqnarray}

Rewriting the expression for the partition function as
\begin{eqnarray}
Z=Z_b\,Z_f\left(1+2\,e^{-\frac{
\beta}{2}(\omega_f+\omega_b)}\,\frac{Z_b}{Z_f}
\frac{
      \beta\,{\rm sh}\left(-\frac{\beta}{2}|\omega_b-\omega_f|\right)
     }{|\omega_b-\omega_f|}
|g|^2+\cdots\right)
\end{eqnarray}
and assuming that $\omega_f\sim M_{Pl}$, the effective correction 
to the distribution function is
\begin{eqnarray}
\Omega=\Omega_0-\frac{e^{-\beta\omega_b}\,Z_b}{(M_{Pl}-\omega_b)}|g|^2+
\cdots= \Omega_0 -\frac{n_b(\omega_b)}{(M_{Pl}-\omega_b)}|g|^2+\cdots 
\nonumber\\
n(\omega)-n_0(\omega)=-n_0(\omega)\frac{|g|^2}{M_{Pl}^2}+
\frac{e^{\beta\omega}}{(e^{\beta\omega}-1)^2}{\beta|g|^2\over M_{Pl}}.
\label{distribution1}
\end{eqnarray}
Here $n_0$ is the free bosonic distribution function.


Consider next the case of two bosons coupled 
in the same way as in the 
Hamiltonian (\ref{hamiltonian}):
\begin{equation}
H=\omega_{1}a^\dagger_1 a_1 + \omega_{2}a_2^\dagger a_2 + g a_1^\dagger a_2 +
g^* a^\dagger_2 a_1
\label{hamiltonian2}
\end{equation}
Here both $a_1$ and $a_2$ are bosons.
Contrary to the previous case this model is exactly solvable and
the spectrum is 
\begin{eqnarray}
E_{n_1,n_2}&=&\omega_{+}n_1+\omega_{-}n_2,\ \ \ n_1,n_2\in {N},\\
\omega_{\pm}&=&\frac{1}{2}\{(\omega_1+\omega_2)\pm
\sqrt{(\omega_1-\omega_2)^2+4g^2}\}\nonumber,\\  
Z&=&\frac{1}{1-e^{-\omega^{+}\beta}}\frac{1}{1-e^{-\omega^{-}\beta}}.
\nonumber
\end{eqnarray}
Here we take $\omega_1=\omega$ to be the 
``photon'' frequency  and $\omega_2$  to be that of the mixing bosonic mode. 
The distribution function for the photon is
\begin{equation}
n(\omega)=\frac{1}{e^{\beta\omega_{-}}-1}\frac{\partial\omega_{-}}
{\partial\omega}+
\frac{1}{e^{\beta\omega_{+}}-1}\frac{\partial\omega_{+}}
{\partial\omega} 
\label{bosonboson}
\end{equation}
If one expands this expression in coupling constant $g$
in the limit $\omega=\omega_1<<\omega_2\sim M_{Pl}$,
the result is identical to that of
(\ref{distribution1}) to order 
$|g|^2$, with the same ``grey body'' factor
and correction to the low energy behaviour.
Unfortunately, the difference
between (\ref{distribution1}) and (\ref{bosonboson}), therefore, 
comes only in the next order of perturbation theory. 
After some algebra one finds
\begin{eqnarray}
&&\Delta n(\omega)_{b+g}=\left\{
 -\frac{1}{M_{Pl}^2}\frac{1}{(e^{\beta\omega}-1)}+
\frac{1}{M_{Pl}}
\frac{\beta\,e^{\beta\omega}}{(e^{\beta\omega}-1)^2}
\right\}g^2+\nonumber\\
&&+\left\{
\frac{\beta^2}{M^2_{Pl}}\left(\frac{e^{2\beta\omega}+e^{\beta\omega}}
{(e^{\beta\omega}-1)^3}\right)+
\frac{\beta}{M^3_{Pl}}\left(\frac{2-4e^{2\beta\omega}-6e^{\beta\omega}}
{(e^{\beta\omega}-1)^3}\right)+
\frac{6}{M^4_{Pl}}\left(\frac{e^{\beta\omega}+1}{(e^{\beta\omega}-1)^2}\right)
\right\}g^4+\cdots\nonumber\\
&&\Delta n(\omega)_{b+b}=\left\{
 -\frac{1}{M_{Pl}^2}\frac{1}{(e^{\beta\omega}-1)}+
\frac{1}{M_{Pl}}
\frac{\beta\,e^{\beta\omega}}{(e^{\beta\omega}-1)^2}
\right\}g^2+\nonumber\\
&&+\left\{
\frac{\beta^2}{M^2_{Pl}}\left(\frac{e^{2\beta\omega}+e^{\beta\omega}}
{(e^{\beta\omega}-1)^3}\right)+
\frac{\beta}{M^3_{Pl}}\left(\frac{-4e^{2\beta\omega}}
{(e^{\beta\omega}-1)^3}\right)+
\frac{6}{M^4_{Pl}}\left(\frac{1}{(e^{\beta\omega}-1)}
\right)
\right\}g^4+\cdots
\label{secondorder}
\end{eqnarray}
where $\Delta n(\omega)_{b+g}$ is the left-hand side of (\ref{distribution1})
while $\Delta n(\omega)_{b+b}$ is the difference of 
(\ref{bosonboson}) and $n_0(\omega)$.
These corrections become identical in the ultraviolet limit
$\beta\omega\rightarrow \infty$ and the only difference
comes in the subleading order in $1/M_{Pl}$.


\end{document}